\documentclass[conference]{IEEEtran}
\IEEEoverridecommandlockouts

\usepackage{cite}
\usepackage{amsmath,amssymb,amsfonts}
\usepackage{graphicx}
\usepackage{textcomp}
\usepackage{xcolor}
\def\BibTeX{{\rm B\kern-.05em{\sc i\kern-.025em b}\kern-.08em
    T\kern-.1667em\lower.7ex\hbox{E}\kern-.125emX}}
    \usepackage{pgfplots}  
\pgfplotsset{width=6.6cm,compat=1.7}
\usepackage{caption}
\usepackage{graphicx}%
\usepackage{multirow}%
\usepackage{amsmath,amssymb,amsfonts}%
\usepackage{amsthm}%
\usepackage{mathrsfs}%
\usepackage{xcolor}%
\usepackage{textcomp}%
\usepackage{manyfoot}%
\usepackage{booktabs}%
\usepackage{algorithm}%
\usepackage{algorithmicx}%
\usepackage{algpseudocode}%
\usepackage{listings}%
\begin{document}

\title{An Explainable AI Model for the Detecting Malicious Smart Contracts Based on EVM Opcode Based Features\\
{\footnotesize \textsuperscript{*}Note: Sub-titles are not captured for https://ieeexplore.ieee.org  and
should not be used}
\thanks{Identify applicable funding agency here. If none, delete this.}
}

\author{\IEEEauthorblockN{1\textsuperscript{st} Roopak Surendran}
\IEEEauthorblockA{\textit{Security Researcher} \\
}}

\maketitle

\begin{abstract}
Hackers may create malicious solidity programs and deploy it in the Ethereum block chain. These malicious smart contracts try to attack legitimate programs by exploiting its vulnerabilities such as reentrancy, tx.origin attack, bad randomness, deligatecall and so on. This may lead to drain of the funds, denial of service and so on . Hence, it is necessary to identify and prevent the malicious smart contract before deploying it into the blockchain. In this paper, we propose an ML based malicious smart contract detection mechanism by analyzing the EVM opcodes. After balancing the opcode frequency dataset with SMOTE algorithm, we transformed opcode frequencies to the binary values (0,1) using an entropy based supervised binning method. Then, an explainable AI model is trained with the proposed binary opcode based features. From the implementations, we found that the proposed mechanism can detect 99\% of malicious smart contracts with a false positive rate of only 0.01. Finally, we incorporated LIME algorithm in our classifier to justify its predictions. We found that, LIME algorithm can explain why a particular smart contract app is declared as malicious by our ML classifier based on the binary value of EVM opcodes. 

\end{abstract}

\begin{IEEEkeywords}
component, formatting, style, styling, insert.
\end{IEEEkeywords}

\section{Introduction}
Smart contracts are the programs deployed in a block chain network that runs only when some predefined conditions or rules met \cite{mohanta2018overview,zheng2017overview}. It enables to automate the execution of the business transactions without the involvement of a third party. Solidity is one of the most popular language for implementing smart contracts which can be deployed in the Ethereum \cite{khan2020ethereum}, a decentralized block chain network.  Solidity is an object-oriented programming language adopted from the concepts of JavaScript, C++ and python \cite{wohrer2018smart}. The programs developed using solidity language run in a virtual environment called EVM (Ethereum Virtual Machine).
\par
Like other programming languages, security vulnerabilities may occur in solidity programming language also \cite{khan2020ethereum}.  An attacker can create and deploy a malicious contract that exploits the vulnerabilities in another deployed legitimate contacts.  This might lead to the attacks such as drainage of funds, DoS/DDoS and so on. Hence, it is required to detect the malicious contract and block it before deploying it into the block chain network.  
There are more than ten kinds of vulnerabilities existing in a solidity program. Some of the major vulnerabilities are:
\begin{itemize}
\item{Bad Randomness: With this vulnerability, an attacker can predict the output of solidity program functions that relies on randomness;}
\item{Denial of Service:  A smart contract can deny the functions of other legitimate contracts by performing DoS attack.;}
\item{Tx.Origin Attack: It is a kind of phishing attack which deceives the contract owner to perform the privileged operations;}
\item{Reentrancy Attack:  This is a recursive process which allows unauthorized transfer of funds from the contract of legitimate user to the malicious contract;}
\item{Forced Ether Reception: With this vulnerability, an attacker can forcefully send ether/s to any other vulnerable contracts;} 
\item{Hiding Malicious Code: This vulnerability allows to hide a malicious code inside another contact.}
\end{itemize}
\par
There are several tools such as mythrill, slither and so on are introduced for analyzing vulnerabilities in a solidity program \cite{feist2019slither,di2019survey}.   These existing tools can only identify known vulnerabilities in a solidity application. It cannot discover unknown or zero-day vulnerabilities in an application. In the proposed mechanism, we use ML algorithms to identify whether a smart contract program tries to attack other applications or not. Further, the ML classifier can justify its decision using explainable AI algorithms. 
\par
Existing mechanisms uses solidity program code or EVM opcode-based features for identifying the vulnerabilities. While deploying the solidity program in EVM, it is compiled to the low level program called EVM byte code. After deploying, a developers can remove their actual high level solidity program from the Etherium network. Hence, the solidity program codes are not always available for analysis.  Hence, in this paper, we used EVM opcode-based features for identifying whether a contact is malicious or not. 
\par
In our collected smart contract dataset, the percentage of malicious contacts are below 1\%. That is, the dataset is highly imbalanced.   Training an ML classifier with a highly imbalanced dataset may lead to the biased prediction towards the majority class.  In order to overcome this limitation, we use data oversampling techniques such as SMOTE to balance the dataset by adding more malicious contracts. After balancing the opcode frequency dataset with SMOTE algorithm, we transformed opcode frequencies to the binary values (0 or 1) using an entropy based supervised binning method. Then, an explainable AI model is trained with the proposed binary opcode based features.  From the implementations, we get 0.99 recall in detecting malicious contract with a false positive rate of only less than 1\%. Further, we used explainable AI algorithm such as LIME to justify the predictions of the sample malicious and legitimate contract.  We found that, our model can explain its predictions based on the values of key EVM opcode level features.
\par
The rest of the paper is organized as follows. In Section 2, literature review has been given. In Section 3, our proposed mechanism has been given. The results and discussions are given in Section 4. In Section 5, we conclude our research and future directions are given.
\section{Literature Review}
In this section, we discuss about the existing solidity vulnerability analyzers in detail.  The existing vulnerability analyzers inspect either the solidity program code or EVM opcode (or a combination of both) for detecting the vulnerabilities.  
\par
 In Contractward \cite{wang2020contractward}, the relationship among solidity code, opcode bigram features are used for identifying the vulnerabilities. In Eth2Vec \cite{ashizawa2021eth2vec}, evm opcodes are used as features of ML classifier to identify whether a contract is vulnerable or not. Grieco et al. \cite{grieco2020echidna} developed a tool called Echidna to detect vulnerabilities by analyzing the solidity code.  Ethainter \cite{brent2020ethainter} can detect composite vulnerabilities by checking information flow with data sanitization in solidity code. Ethver \cite{mazurek2021ethver}, a formal verification-based vulnerability analyzer which takes solidity codes as input and translate it into MDP (Markov Decision Process) models. Then, it is verified using PRISM model checker. Yang et al. \cite{yang2019fether} proposed a mechanism called FEther for identifying smart contract vulnerabilities by combining symbolic analysis and higher order  logic theorem-proving. In GasGuage \cite{nassirzadeh2022gas}, fuzz testing is used to detect Ethereum gas related vulnerabilities. In Musc \cite{barboni2021sumo}, the mutants of smart contacts are generated from AST (Abstract Syntax Tree) and these AST generated mutants are converted to solidity code for testing.  Gupta et al. \cite{gupta2022deep}, suggested a deep learning based mechanism to detect vulnerable contracts from the EVM opcodes.
 \par
All of the existing mechanisms discussed the problem of detecting vulnerabilities in smart contracts. They did not make an attempt to detect the malicious smart contract which tries to exploit vulnerabilities. Moreover, the classifier decisions has not been justified in all these methods. In order to overcome this limitation, in this work, we propose a mechanism to transform opcode frequency based features to binary features for interpreting classifier decisions using explainable AI algorithm.  From the implementations, we found that our model can effectively interpret why a particular smart contract is declared as malicious by analyzing binary value of EVM opcodes. 

\section{Methodology}
In this section, we discuss about our malicious contract detection mechanism in detail.  The proposed mechanisms consist of three steps. In the first step, we extract EVM opcodes of malware and goodware apps in our dataset. The percentage of malicious contacts may be very lesser than the legitimate contracts. Hence, in the second step, we use oversampling technique such as SMOTE, to balance the dataset by adding more malicious contracts. Then, we train an ML classifier and tested with the features of unknown smart contracts. Further, explainable AI is used to justify the classifier predictions. The architecture of our mechanism is given in Figure 1. 
\begin{figure*}[h!]
        \centering
       
		\includegraphics[scale=.07]{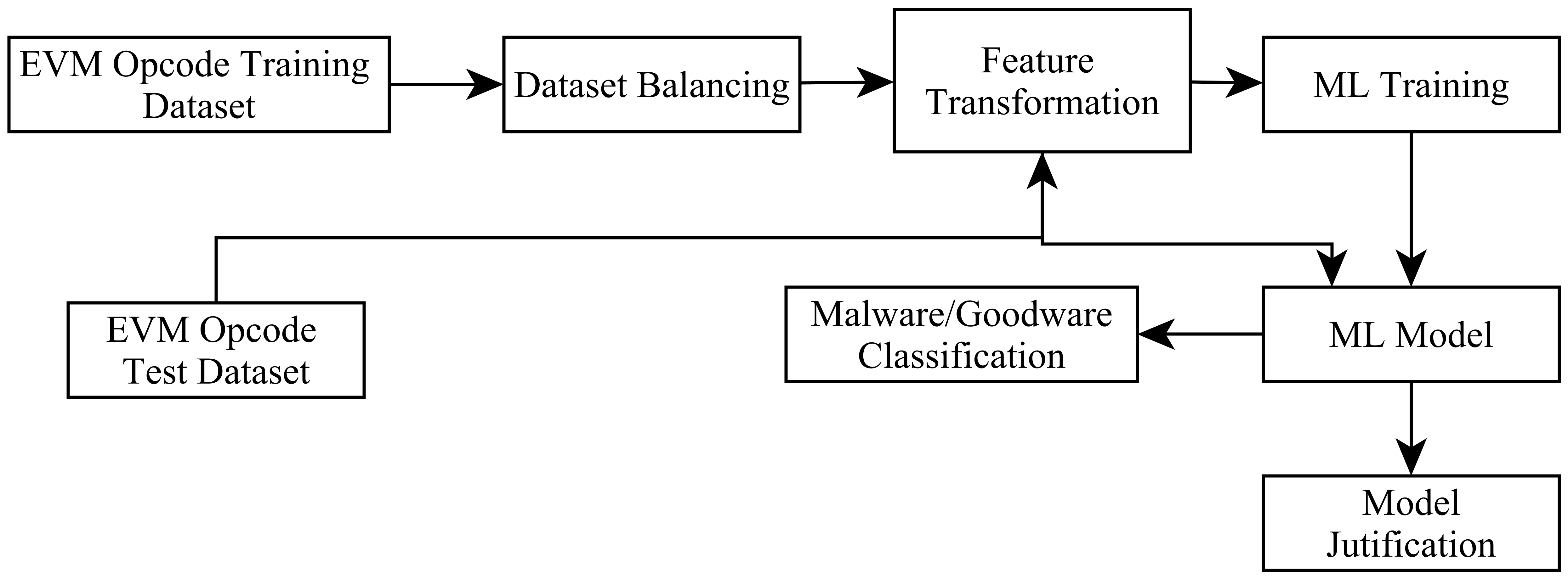}
	 \caption{Proposed Malicious Contract Detection Mechanism}	
\end{figure*}

\subsection{Dataset Oversampling Phase}
The total number of malicious contracts is very lower as compared to benign contracts.  That is, there are only below 1\% of malicious contracts in a dataset.  Training with a highly imbalanced dataset may lead to the performance bias towards the majority class. Hence, in order to balance the training dataset, we used SMOTE algorithm \cite{kotsiantis2006handling}.  The steps in data oversampling using SMOTE algorithm are given below.
\begin{enumerate}
\item{Step 1: Extract the frequency values of $n$ EVM opcodes $X_i$ for $i=1,2,..,n$ in the malware and legitimate contracts in the dataset.}
\item{Step 2: Build a minority class set $X$, for each $x$ $\in$ $X$, the nearest neighbor of $X$ is calculated using Euclidean distance formula. The Euclidean distance $D(A,B)$ of any given two points $A=(a_1,a_2,\dots,a_n)$ and $B=(b_1,b_2,\dots,b_n)$ is calculated as, \begin{equation*}D(A,B)=\sqrt{\sum_{i=1}^{n}(a_i-b_i)}.\end{equation*}
} 
\item{Step 3:  Set the sampling rate $N$ according to the imbalanced population,  for each $x$ $\in$ $X$, $N$ samples are selected for k nearest neighbors to construct a new set $Y$.}
\item{Step 4: For each sample $y_{k}$ $\in$ $Y$ for $k=1,2,…,N$, new samples are generated by the following formula $X+t.|x-y_k|$, where $0 \le t \le 1$.}
\end{enumerate}
\subsection{Feature Transformation Using Entropy Based Supervised Binning Algorithm}
In both training and test dataset, we use binning technique to transform discrete frequency based opcode features to binary categorical features. Assume that, you have a train or test dataset with a continuous feature $X=\{X_i:i=1,2,3,\dots,n\}$ and a corresponding target variable $Y_i \in \{0,1\}$ denotes its label (1 represents malicious and 0 represents legitimate). Entropy is a measure of uncertainty or disorder in a set of data. For a given set of data $S$ with $k$ classes, the entropy $H(S)$ is calculated as:
\begin{equation*}
H(S) = -\sum_{i=1}^{k} p_i \cdot \log_2(p_i),
\end{equation*}
where $p_i$ is the proportion of data points in class $i$. Information Gain is used to measure the reduction in entropy achieved by partitioning a dataset based on a certain attribute (e.g., a continuous feature). The information gain is calculated as:
\begin{equation*}
\text{Information Gain} = H(S) - \sum_{j} \frac{|S_j|}{|S|} \cdot H(S_j),
\end{equation*}
where $S_j$ is the subset of $S$ for which feature $X$ falls into bin $j$. The supervised binning algorithm is given below:
\begin{itemize}
\item{Start by sorting the feature $X$ in ascending order;}
\item{Calculate information gain for each possible split point;}
\item{Choose the split point that maximizes the information gain, effectively dividing the data into two bins;}
\end{itemize}
\par
After identifying the splitting points $S(X_i)$ to the $n$ frequency based opcode features $X_i$ for $i=1,2,3,\dots,n$ in the dataset, we transform opcode frequencies $O(X_i)$ of $X_i$ to binary features $B(X_i)$ by comparing with the corresponding splitting points $S(X_i)$. That is, 
\begin{equation}
B(X_i)=
\begin{cases}
1& \text{if  } O(X_i) \ge S(X_i); \\
0& \text{otherwise}
\end{cases}
\end{equation}

\subsection{Interpreting Blackbox ML Models Using LIME Algorithm}
In this section, we discuss about machine learning based malicious smart contract detection from opcode sequences. After constructing a balanced data set of binary opcode based features, we train an ML classifier. This ML classifier can detect whether an unknown contract is malicious or not.  We incorporate explainable AI algorithm such as LIME \cite{gramegna2021shap} to justify the classifier predictions.  The explanation $E(x)$ of a feature vector $x$ produced by LIME is following:
\begin{equation*}
E(x)=\arg\min_{g \in G} L(f,g,\pi_x)+\Omega(g),
\end{equation*}
where $f$ is the actual model, $g$ is the linear explanation model, $\pi_x$ is the proximity measure between an instance of $z$ to $x$ and $\Omega(g)$ is the model complexity. $\pi_x$ gives the weights $w$ to the perturbed instances $z^{\prime}$ based on their distance from $x$. $L(f,g,\pi_x)$ is calculated as,
\begin{equation*}
L(f,g,\pi_x)=\sum_{z,z^{\prime} \in Z}\pi_x(z)(f(z)-g(z^\prime))^2,
\end{equation*}
where $\pi_x(z)=e^{(-D(x,z)^2/(\sigma^2))}$, $g(z^\prime)=w.g(z)$ and $D(x,z)$ is the distance function.
\section{Results and Discussions}
In this section, we discuss the performance results of our ML based malicious smart contract detection mechanism and their explainability.  We collected the opcode sequences of malicious and benign contracts from the forta dataset \footnote{https://huggingface.co/datasets/forta/malicious-smart-contract-dataset}. The collected dataset is highly imbalanced because there exists only a very few malicious contracts in the wild. The collected dataset consists of 139451 legitimate and 143 malicious contracts. In their mechanism, a logistic regression (LR) classifier was trained with the previously detected malicious opcode patterns of to detect malicious behavior in unseen smart contracts. Because of the high dataset imbalance, the classifier can detect only 59\% of malicious smart contracts. 
\par
The proposed mechanism was implemented in a windows 11 PC with 128GB of memory. After collecting the dataset, we calculated the frequencies of opcodes in each malicious and legitimate contracts. The, we add a label `1' at the end of malicious contract and `0' at the end of legitimate contract. Finally, a CSV dataset file contain malicious and legitimate smart contract opcode frequency based feature vectors was created. The opcodes found in the smart contact programs are given in Table 1. 
\begin{table*}[h!]
\centering
\resizebox{1\textwidth}{!}{
\begin{tabular}{ | l | l |  l | l|}
 \hline
 SELFBALANCE & BASEFEE & MLOAD & MSTORE  \\ \hline 
 SLOAD & SSTORE & JUMP & JUMPI \\ \hline
 PC &  MSIZE & GAS & RETURNDATACOPY  \\ \hline
 EXTCODEHASH & BLOCKHASH&COINBASE & TIMESTAMP \\ \hline 
NUMBER &   PREVRANDAO & GASLIMIT & CHAINID \\ \hline
BALANCE & ORIGIN & CALLER & CALLVALUE \\ \hline
CALLDATALOAD & CALLDATACOPY & CODESIZE & CODECOPY 
\\ \hline
GASPRICE&EXTCODESIZE&EXTCODECOPY& SLT \\ \hline
SGT&EQ&ISZERO & AND \\ \hline
OR&XOR&NOT&BYTE \\ \hline
SHL&SHR&SAR& KECCAK256 \\ \hline
ADDRESS&CALLCODE&RETURN&DELEGATECALL \\ \hline
CREATE&STATICCALL & REVERT&SELFDESTRUCT \\ \hline
LOG&SWAP&DUP&PUSH  \\ \hline
POP&STOP&ADD&MUL \\ \hline
SUB&CALL&DIV&SDIV \\ \hline
MOD&SMOD&ADDMOD&MULMOD \\ \hline
EXP&SIGNEXTEND&LT&GT \\ \hline
CREATE&&& \\ \hline

\end{tabular}}
\caption{List of Opcodes in Solidity Program}
\end{table*}
\par
It is known that, an ML model will be biased to the features of majority class if we train it with a highly imbalanced dataset.  Initially, we transformed  frequency based opcode features to binary features using the entropy based supervised binning algorithm.  Then, we used 70\% of malicious smart contract (93 apps) and 138247 legitimate contract for training ML classifier. The training dataset has been balanced by adding equal number of synthetically generated malicious contract features using SMOTE algorithm. Then, we train an ML classifier in the SMOTE balanced dataset of transformed binary opcode based features corresponding to 70\%  of malicious smart contracts and tested those of remaining 30\% of malicious smart contracts for evaluation. The classifiers \cite{kotsiantis2007supervised} used for evaluation are:
\begin{itemize}
    \item{Naive Bayes;}
    \item{Logistic Regression;}
    \item{K- Nearest Neighbour;}
    \item{Decision Tree.}
\end{itemize}
\begin{table*}[h!]
\centering
\begin{tabular}{|l|l|l|l|l|l|}
\hline
Algorithm & TPR & FPR & Precision & Accuracy & F1Score \\ \hline
Naive Bayes & 0.96 & 0.04 & 0.99 & 0.99  & 0.98 \\ \hline
LR & 0.99 & 0.01& 0.99& 0.99& 0.99\\ \hline
DT & 0.99 & 0.01& 0.99& 0.99& 0.99\\ \hline
KNN & 0.90 & 0.10& 0.92& 0.99& 0.91\\ \hline
\end{tabular}
\caption{Performance of ML Classifiers in Our Dataset}
\end{table*}
The performance results in different machine learning classifiers are given in Table 2.  From the Table 2, we can see that the proposed mechanism can detect 99\% of malicious smart contracts with a false positive rate of only below 1\%. The comparison results against the mechanism used in Forta is given in Table 3. From Table 3, we can see that our mechanism outperforms Forta with a recall rate of 0.99. In order to find out the relevant opcode based features, we used extra trees feature selection method in the opcode frequency dataset \cite{kharwar2022ensemble}.  The relevant opcode features $X_i$, its frequencies $O(X_i)$, splitting points $S(X_i)$ and assigned binary labels $B(X_i)$ of a sample smart contract application is given in Table 4. In rest of our work, we use only these relevant opcodes for justifying the classifier predictions. 
\begin{table*}[h!]
\centering
\begin{tabular}{|l|l|l|l|l|l|}
\hline
Method & TPR & FPR & Precision & Accuracy & F1Score \\ \hline
Proposed Method & 0.99 & 0.01& 0.99& 0.99& 0.99\\ \hline
Forta & 0.59 & N/A& 0.88& N/A& N/A \\ \hline
\end{tabular}
\caption{Performance of Our Mechanism against Forta}
\end{table*}

\begin{table*}[h!]
\centering
 \resizebox{\textwidth}{!}{  

    \begin{tabular}{|l|l|l|l|}
    \hline
    EVM Opcode & Opcode Frequency & Splitting Point & Binary Value \\ \hline
 SSTORE & 10 & 17 & 0 \\ \hline
 RETURNDATACOPY & 1 & 17 & 0 \\ \hline
 SLT &0 &0 & 0 \\ \hline
 EQ &16& 32 & 0 \\ \hline 
 OR &129 &163 & 0 \\ \hline
 RETURN & 17 & 75 &0  \\ \hline
 DELEGATECALL &1 & 4 & 0 \\ \hline
LOG & 3 & 5 & 0  \\ \hline
 SUB & 57 & 37 & 1 \\ \hline
 SLOAD & 21 & 63 & 0 \\ \hline
    \end{tabular}}
    \caption{Details of Key EVM Opcodes and Conditions for Binary Feature Transformation}
\end{table*}
\par
\subsection{Illustrating Explainability of ML with a Malicious and Legitimate Contract }
In this section, we illustrate the explainability of our ML classifier in a sample malicious and legitimate contract. We fed the binary opcode based features of the malicious and legitimate contracts in a trained explainable AI model. Here, we made justification on the basis of top features those have a contribution value greater than zero. The classifier justification for a malicious contract is given in Table 5. From Table 5, we can see that, the majority of features supports malicious class label. In Figure 2, we can see that the total contributions of features support malicious contract is higher than that of legitimate. Hence, it is clear that the features corresponding to the smart contract is malicious. The classifier justification for legitimate contract is given in Table 6. From Table 6, we can see that, the majority of features supports legitimate class label. In Figure 3, we can see that the total contributions of features support legitimate contract is higher than that of malicious.  Hence, it is clear that the features corresponding to the smart contract is legitimate. 
\begin{table*}
 \resizebox{\textwidth}{!}{  
    \begin{tabular}{|l|l|l|l|}
    \hline
    Feature & Actual Label Value & Supported Class Label & Contribution \\ \hline
    RETURNDATACOPY & 1 & Malicious & 0.25 \\ \hline
    RETURN & 1 & Malicious & 0.21 \\ \hline
    SLT & 1 & Malicious & 0.21 \\ \hline
    EQ & 1 & Legitimate & 0.19 \\ \hline
    LOG & 1 & Legitimate & 0.18 \\ \hline
    DELIGATECALL & 0 & Malicious & 16 \\ \hline
    OR & 0 & Malicious & 10 \\ \hline
    SSTORE & 0 &Malicious & 0.03 \\ \hline
    SLOAD & 0 & Malicious & 0.03 \\ \hline
    SUB & 0 & Malicious & 0.03 \\ \hline
       \end{tabular}}
    \caption{LIME Algorithm Based Justification for Classifier Prediction in Malicious Contract}
\end{table*}
\pgfplotsset{every tick label/.append style={font=\small}}

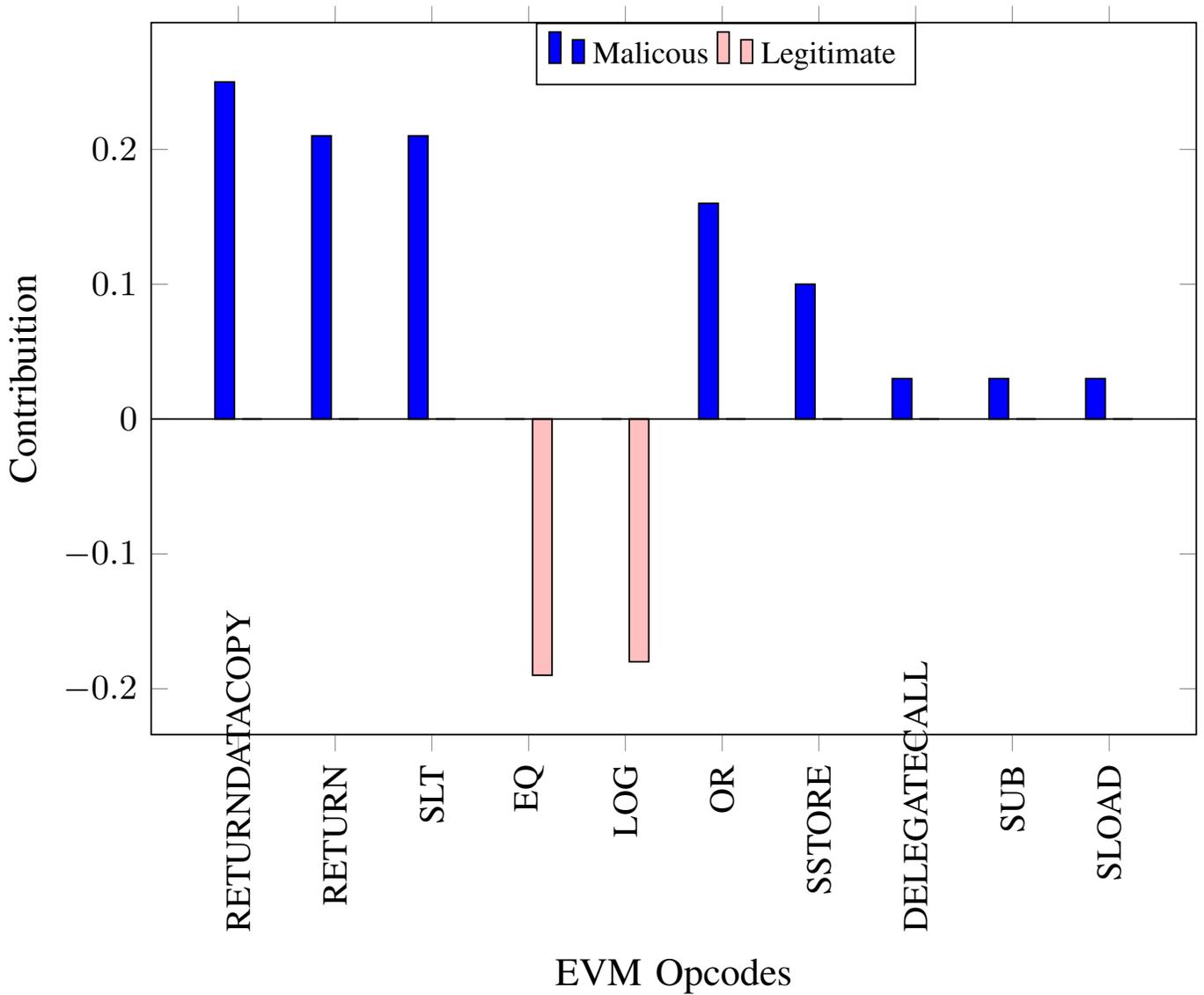
\begin{figure*}[h!]
\centering
\resizebox{\textwidth}{!}{%
\begin{tikzpicture}
    \begin{axis}[
        ybar,
        width=11cm,
        height=8cm,
        legend style={
            area legend,
            at={(0.55,1)},
            anchor=north,
            legend columns=-1,
            font=\footnotesize,     
        },
        xtick=data,
        symbolic x coords={
            RETURNDATACOPY,
            RETURN,
            SLT,
            EQ,
            LOG,
            OR,
            SSTORE,
            DELEGATECALL,
            SUB,
            SLOAD
        },
        x tick label style={
            rotate=90,
            anchor=east,
            align=right,
            text width=1.5cm,
        },
        xlabel={EVM Opcodes},
        ylabel={Contribuition},
        bar width=5pt,
    ]
        \addplot [black,fill=blue] coordinates {
            (RETURNDATACOPY,0.25)
            (RETURN,0.21)
            (SLT,0.21)
            (EQ,0)
            (LOG,0)
             (OR,0.16)
            (SSTORE,0.10)
            (DELEGATECALL,0.03)
            (SUB,0.03)
            (SLOAD,0.03)
        };
        \addplot [black,fill=pink] coordinates {
            (RETURNDATACOPY,0)
            (RETURN,0)
            (SLT,0)
            (EQ,-0.19)
            (LOG,-0.18)
            (OR,0)
            (SSTORE,0)
            (DELEGATECALL,0)
            (SUB,0)
            (SLOAD,0)

        };
        \draw (axis cs:{[normalized]\pgfkeysvalueof{/pgfplots/xmin}},0)
            -- (axis cs:{[normalized]\pgfkeysvalueof{/pgfplots/xmax}},0);

        \legend{
            Malicous,
            Legitimate,
            PL2,
        }
    \end{axis}
\end{tikzpicture}}
\caption{Contributions of EVM Opcode Features in Malicious Contract}
\end{figure*}
\begin{table*}
 \resizebox{\textwidth}{!}{  
    \begin{tabular}{|l|l|l|l|}
    \hline
    Feature & Actual Label Value & Supported Class Label & Contribution \\ \hline
    RETURNDATACOPY & 0 & Legitimate & 0.24 \\ \hline
    RETURN & 0 & Legitimate & 0.21 \\ \hline
    SLT & 1 & Malicious & 0.20 \\ \hline
    EQ & 0 & Malicious & 0.18 \\ \hline
    LOG & 1 & Legitimate & 0.16 \\ \hline
    OR & 1 & Legitimate & 0.08 \\ \hline
    SSTORE & 1 & Legitimate & 0.08 \\ \hline
    DELIGATECALL & 0 &  Malicious & 0.08 \\ \hline
    SUB & 1 & Malicious & 0.03 \\ \hline
    SLOAD & 1 & Legitimate & 0.02 \\ \hline
    \end{tabular}}
    \caption{LIME Algorithm Based Justification for Classifier Prediction in Legitimate Contract}
\end{table*}
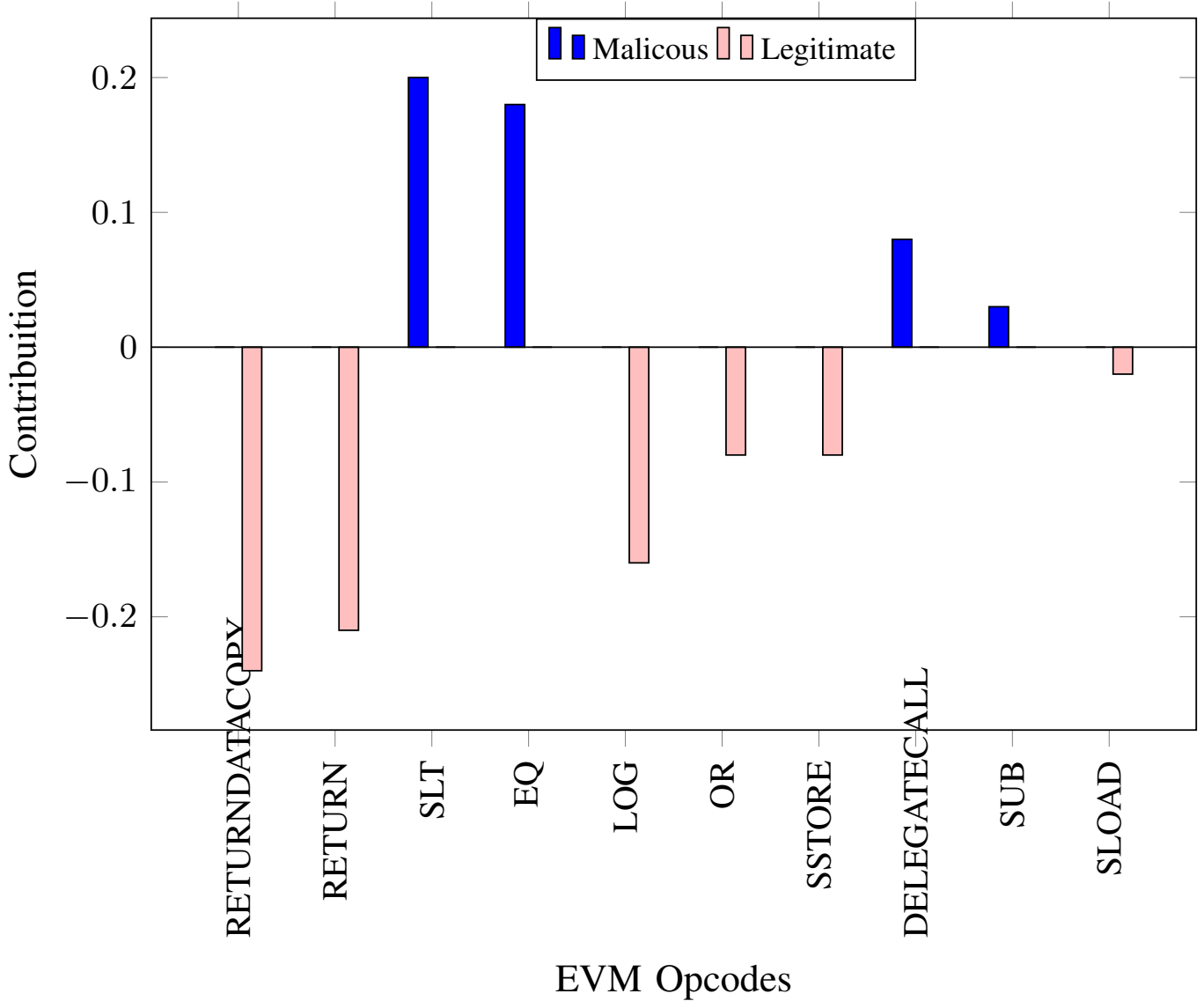
\begin{figure*}[h!]
\centering
\resizebox{\textwidth}{!}{%
\begin{tikzpicture}
    \begin{axis}[
        ybar,
        width=11cm,
        height=8cm,
        legend style={
            area legend,
            at={(0.55,1)},
            anchor=north,
            legend columns=-1,
            font=\footnotesize,     
        },
        xtick=data,
        symbolic x coords={
            RETURNDATACOPY,
            RETURN,
            SLT,
            EQ,
            LOG,
            OR,
            SSTORE,
            DELEGATECALL,
            SUB,
            SLOAD
        },
        x tick label style={
            rotate=90,
            anchor=east,
            align=right,
            text width=1.6cm,
        },
        xlabel={EVM Opcodes},
        ylabel={Contribuition},
        bar width=5pt,
    ]
        \addplot [black,fill=blue] coordinates {
            (RETURNDATACOPY,0)
            (RETURN,0)
            (SLT,0.20)
            (EQ,0.18)
            (LOG,0)
             (OR,0)
            (SSTORE,0)
            (DELEGATECALL,0.08)
            (SUB,0.03)
            (SLOAD,0)
        };
        \addplot [black,fill=pink] coordinates {
            (RETURNDATACOPY,-0.24)
            (RETURN,-0.21)
            (SLT,0)
            (EQ,0)
            (LOG,-0.16)
            (OR,-0.08)
            (SSTORE,-0.08)
            (DELEGATECALL,0)
            (SUB,0)
            (SLOAD,-0.02)

        };
        \draw (axis cs:{[normalized]\pgfkeysvalueof{/pgfplots/xmin}},0)
            -- (axis cs:{[normalized]\pgfkeysvalueof{/pgfplots/xmax}},0);

        \legend{
            Malicous,
            Legitimate,
            PL2,
        }
    \end{axis}
\end{tikzpicture}}
\caption{Contributions of EVM Opcode Features in Legitimate Contract}

\end{figure*}

\section{Conclusion and Future Scope}
In this section, we discuss about the limitations and future directions of our work. Here, we build an explainable ML classifier which is capable to detect malicious contracts and make justifications on its prediction. Our ML classifier can able to detect malicious smart contracts with a precision and F1 score of 0.99. With the help of explainable AI algorithm, we found that the frequencies of certain opcodes are very high in malicious contracts. 
\par
It is possible for an adversary to defeat our ML classifier using adversarial attacks \cite{ibitoye2019threat}. In future, we will build an adversarial resistant ML classifier to detect adversarial malicious contracts. For that, we will synthetically generate malicious contracts similar to the real world scenario using conditional GAN and train the adversarial resistant classifier with it \cite{xu2019modeling}. Hence, in future, the proposed mechanism will be capable to detect unknown adversarial malicious contract attacks.
\par
Our mechanism justifies the predictions on the basis of frequency values of certain opcodes.  With respect to the change in evolving malicious contracts, these frequency range value of malicious and legitimate contracts may change in future. In this situtation, explainable AI algorithm will not work \cite{de2022perils}. In order to overcome this limitations, we will combine semantic analysis algorithms with explainable AI to make more reliable explanations for evolving smart contracts in future. 
\bibliographystyle{elsarticle-num}  \bibliography{bibiliography}






\end{document}